\documentclass[twocolumn,aps,showpacs,amsmath,amssymb,prb,amsmath,amssymb]{revtex4}
\usepackage{hyperref}
\usepackage{dcolumn}
\usepackage{amssymb}
\usepackage{graphicx,xcolor}
\usepackage{hyperref}
\usepackage{soul}

\newcommand{\vvec}[1]{\textbf{\textit{#1}}}
\newcommand{\gradi}[0]{^{\circ}}
\newcommand{\mmu}[0]{\mu\mathrm{m}}

\begin{document}

\title{Symmetric polarization insensitive directional couplers fabricated by femtosecond laser waveguide writing}

\author{Giacomo Corrielli$^{1,2}$}
\email{giacomo.corrielli@mail.polimi.it}
\author{Simone Atzeni$^{1,2}$}
\author{Simone Piacentini$^{2}$}
\author{Ioannis Pitsios$^{1}$}
\author{Andrea Crespi$^{1,2}$}
\author{Roberto Osellame$^{1,2}$}

\affiliation{$^1$Istituto di Fotonica e Nanotecnologie - Consiglio Nazionale delle Ricerche (IFN-CNR), piazza Leonardo da Vinci 32, 20133 Milano, Italy}
\affiliation{$^2$Dipartimento di Fisica, Politecnico di Milano, piazza Leonardo da Vinci 32, 20133 Milano, Italy}

\date{\today}

\begin{abstract}
We study analytically the polarization behaviour of directional couplers composed of birefringent waveguides, showing that they can induce polarization transformations that depend on the specific input-output path considered. On the basis of this study, we propose and demonstrate experimentally, by femtosecond laser writing, directional couplers that are free from this problem and also yield a polarization independent power-splitting ratio. More in detail, we devise two different approaches to realize such devices: the first one is based on local birefringence engineering, while the second one exploits ultra-low birefringence waveguides obtained by thermal annealing.
\end{abstract}


\maketitle
\section{Introduction}\label{intro}

Linear integrated-optical devices\cite{iga2005} are used for splitting, combining or elaborating signals in classical optical communications, and stand at the basis of complex current experiments in quantum photonics\cite{carolan2015, harris2017}. As they rely, for their operation, on field superposition and interference, they often require control on the polarization of the propagating light\cite{dai2012}. This is especially true if information is encoded on such a degree of freedom of the light signal.

In the recent years, femtosecond laser micromachining (FLM) in glass\cite{davis1996, eaton2008} has proved to be a powerful fabrication technique for integrated polarization processing, especially for quantum information applications. On the one hand, the low birefringence of femtosecond laser written waveguides enables the propagation of polarization encoded qubits without coherence degradation \cite{sansoni2010}. On the other hand, the versatility of this technology has allowed 
to exploit such minimal birefringence for the fabrication of devices for polarization manipulation, such as polarization splitters\cite{crespi2011, fernandes2011, dyakonov2017, pitsios2017} and integrated retarders and waveplates \cite{fernandes2011r, corrielli2014, heilmann2014}.

In several applications, however, it is required to achieve a complete polarization insensitivity of the device operation; to this aim, a crucial component to develop is a polarization insensitive directional couplers (PIC). PICs have been demonstrated by FLM employing a unique three-dimensional geometry \cite{sansoni2012}. However, an accurate characterization showed that power splitting is rigorously equal only for horizontally and vertically polarized input states, and small unwanted polarization rotations may anyway occur to light propagating in those devices \cite{vest2015}.  More recently, Pitsios et al. reported PICs operating at telecom wavelengths, fabricated by FLM in borosilicate glass with a fully planar layout\cite{pitsios2017}, which provide equal power splitting for every input polarization state. 
Actually, as we will study in more detail in the following, despite the equal power splitting, all couplers made of birefringent waveguides may produce polarization rotations that depend on the specific input-output path taken by the light into the coupler. This effect may represent a strong limitation to achieve a true polarization independent operation and in developing integrated devices for polarization-state engineering.

In this work we study analytically the polarization behaviour of directional couplers and we propose and validate experimentally two different strategies for realizing PICs free of any asymmetry using the FLM technology.

\section{Theoretical model}\label{theo}
\begin{figure*}[t]
\centering
\includegraphics[]{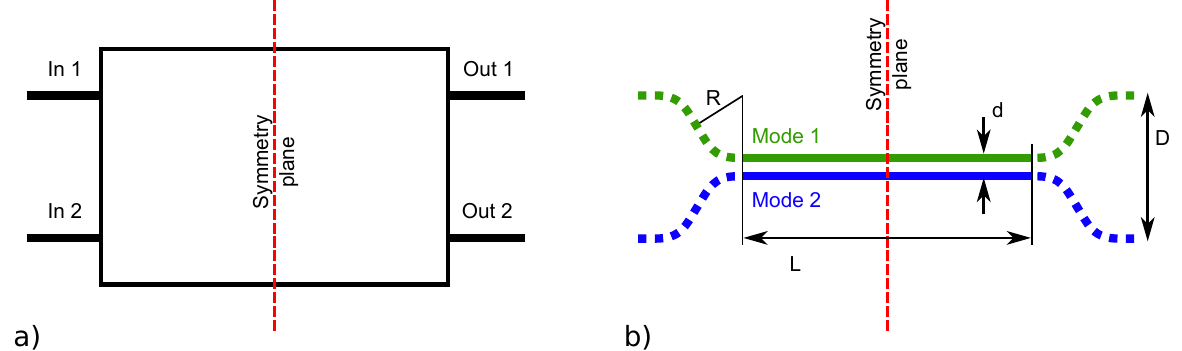}
\caption{(a) Schematic of a generic 2x2 optical device, symmetric under reflection with respect to a plane passing through its center and orthogonal to light propagation direction. (b) Schematic of a directional coupler of length L, composed by two waveguides with different birefringence. d is the coupling distance, D is the separation of the waveguides at the input/output of the device and R is the radius of curvature of the S-bands that compose the fan-in/out regions. All couplers presented in this work have been fabricated with D = 127 $\mmu$ and R = 120 mm.}
\label{f1}
\end{figure*}

\subsection*{Polarization behavior of a unitary and symmetric 2-ports device}
We derive in the following a simple analytical model to describe the operation of a reciprocal linear optical device with two input and two output ports, under the hypotheses that: (I) the device is physically symmetric with respect to a transverse plane as in Fig.~\ref{f1}a; (II) the device supports modes with two well-defined orthogonal polarization states, and orthogonal polarization states do not mix within the device; (III) the device is lossless.

The most general 2$\times$2 unitary matrix $\hat{U}$, which describes the operation of an arbitrary lossless device acting on two optical modes, is written as:
\begin{equation}
\hat{U}= e^{i\phi}\left[
\begin{array}{cc}
r& it \\
it^*& r^*\\
\end{array}
\right],
\label{Generic1}
\end{equation}
where $i$ is the imaginary unit, $r$ and $t$ are complex numbers such that $|r|^2+|t|^2=1$, and $\phi$ is a generic global phase term. 

The symmetry condition (I) is mathematically expressed by the relation $\hat{U}= \hat{U}^\mathrm{T}$\cite{potton2004}. In addition, considering both light polarizations, our two-port device is actually acting on four optical modes. For convenience and with no loss of generality, we shall further detail hypothesis (II) assuming that the device is built of a material or optical components with a fixed birefringence axis direction, oriented either vertically or horizontally: thus, to each spatial mode correspond two distinct polarization modes, vertically (V) or horizontally (H) polarized, which does not exchange power within the device. Applying these constraints, we can write the overall transfer matrix as:
\begin{equation}
\resizebox{.90 \linewidth}{!}
{
$\hat{U}_\mathrm{FULL}= e^{i\phi_H}\left[
\begin{array}{cccc}
r_He^{-i\theta_H} & it_H & 0 & 0 \\
it_H & r_He^{i\theta_H} & 0 & 0 \\
0 & 0 & r_Ve^{i(\Phi - \theta_V)} & it_Ve^{i\Phi} \\
0 & 0 & it_Ve^{i\Phi} & r_Ve^{i(\Phi + \theta_V)} \\
\end{array}
\right],$
}
\label{Overall}
\end{equation}
where we have further defined the phase difference $\Phi = \phi_V-\phi_H$.
Such matrix acts on the vector of the complex electric field amplitudes:
$\vvec{E}=[E_{\mathrm{1H}}, E_{\mathrm{2H}}, E_{\mathrm{1V}}, E_{\mathrm{2V}}]^\mathrm{T}$
with subscripts indicating the spatial mode and the polarization. 
It is worth to note that $\hat{U}_\mathrm{FULL}$ is a symmetric matrix too. 

From (\ref{Overall}) one can extract the matrices $\hat{J}_{m\rightarrow n}$ (analogous to Jones matrices) that define the polarization transformation occurring to light when it travels from input $m$ to output $n$:
\begin{align}
\hat{J}_{1\rightarrow 1} &=
e^{i\phi_H}\left[
\begin{array}{cc}
r_He^{-i\theta_H} & 0 \\
0 & r_Ve^{i(\Phi-\theta_V)}\\ 
\end{array}
\right], \nonumber \\ 
\hat{J}_{1\rightarrow 2} &=
\hat{J}_{2\rightarrow 1} = e^{i\phi_H}\left[
\begin{array}{cc}
it_H & 0 \\
0 & t_Ve^{i\Phi}\\
\end{array}
\right], \nonumber\\
\hat{J}_{2\rightarrow 2} &=
e^{i\phi_H}\left[
\begin{array}{cc}
r_He^{i\theta_H} & 0 \\
0 & r_Ve^{i(\Phi+\theta_V)}\label{Jones1}\\ 
\end{array}
\right]. 
\end{align}
Note that, in general, such transformations can be different for each light path and act both on the phase and on the relative amplitude of the field components. Moreover, despite $\hat{U}_{\mathrm{FULL}}$ is unitary, the $\hat{J}_{m\rightarrow n}$ are not, consistently with the fact that light power is conserved only globally in the device.
 
In the case of a balanced power splitter, in which $r_H = r_V$, the polarization transformations (\ref{Jones1}) are equivalent to those of birefringent waveplates with the fast axis oriented vertically (or horizontally), that just introduce a phase delay between the H and V polarization modes without altering their relative amplitude. For light transmitted on the crossed spatial optical mode (1$\Rightarrow$2 or 2$\Rightarrow$1) such phase delays are identical and equal to $\Phi$. For the light that remains on the same spatial mode, the phase shifts is $\Phi\pm\psi$, where $\psi=\theta_H-\theta_V$, and the $\pm$ signs hold for the paths 1$\Rightarrow$1 and 2$\Rightarrow$2 respectively.

These results are valid for a generic two-port devices fulfilling the initial hypotheses (I), (II), (III), regardless of its internal functioning. The latter determines the specific values assumed by the various terms that appear in (\ref{Overall}). As well, this analysis holds for lossy devices if losses are uniform, provided that the matrices are considered as operating on normalized vectors $\vvec{E}_N=\vvec{E}/|\vvec{E}|$.

\subsection*{Waveguide directional coupler}

We shall now specialize our discussion to the case of a directional coupler. For the sake of simplicity, we neglect in this analysis the fan-in and fan-out regions which are present in any real device, and we focus our attention only on the interaction region. This allows us to consider as uniform all the relevant physical coefficients of the coupler, and to retrive a simple analytic expression for the matrices (\ref{Jones1}).

Let us consider two parallel single-mode optical waveguides that can exchange power via evanescent coupling in a region of length $L$ (see figure \ref{f1}b). The propagation constants of the distinct spatial and polarization modes may be different, but the device keeps the crucial features assumed in the previous discussion: a well defined and fixed (vertical or horizontal) birefringence axis, and reflection symmetry with respect to a central plane. Consequently, its transfer matrix still reads as (\ref{Overall}), and the various quantities can be specified as a function of the coupling coefficients $k_{H,V}$ between same-polarization modes and of the different propagation constants $\beta_{1V}, \beta_{2V}, \beta_{1H}, \beta_{2H}$ (the subscripts indicating the spatial and polarization modes).
Precisely, from standard coupled mode theory \cite{yariv1973} it is not difficult to retrieve:
\begin{align}
t_{H,V} &= \dfrac{k_{H,V}}{\sqrt{k^2_{H,V}+ \frac{\Delta^2_{H,V}}{4}}}\sin\left(\sqrt{k^2_{H,V}+ \frac{\Delta^2_{H,V}}{4}} L\right), \nonumber \\  r_{H,V} &= \sqrt{1 - t_{H,V}^2}, \label{relevant1}\\
\theta_{H,V} &= \arcsin\left(\frac{\Delta_{H,V}}{2k_{H,V}}\frac{t_{H,V}}{r_{H,V}}\right),\nonumber \\ 
\phi_{H} &= \frac{\beta_{H1}+\beta_{H2}}{2}L,\label{relevant3}\\
\Phi &= \left(\frac{b_1+b_2}{2}\right)\frac{2\pi}{\lambda}L, \label{relevant4}
\end{align}
where we have further defined the detunings $\Delta_{H} = \beta_{2H}-\beta_{1H}$,  $\Delta_{V} = \beta_{2V}-\beta_{1V}$ and the birefringences $b_{1}=\frac{\lambda}{2\pi}(\beta_{1V}-\beta_{1H})$, $b_{2}=\frac{\lambda}{2\pi}(\beta_{2V}-\beta_{2H})$ ($\lambda$ being the wavelength in vacuum).

In the special case of a PIC, the definition of such a device implies $t_H=t_V=t$ and $r_H=r_V=r=\sqrt{1-t^2}$. In many practical cases, we can also reasonably assume $k_H \simeq k_V=k$ and $\Delta_{H,V}\ll k_{H,V}$. Using these approximations, one can derive:
\begin{equation}
\psi = \theta_H - \theta_V \simeq \frac{\pi}{\lambda k}\frac{t}{r}\left(b_1-b_2\right).
\label{relevant5}
\end{equation}

The expressions \eqref{relevant4} and \eqref{relevant5} relate in a simple fashion the polarization rotations occurring inside a PIC and the birefringence values $b_1$ and $b_2$ of the isolated waveguides: $\Phi$ is proportional to their average whereas $\Psi$ is proportional to their difference.
The fact that $\Psi$ is in general non vanishing, except for a coupler composed of waveguides with identical birefringence, implies that even a PIC can produce polarization rotations that are path dependent and thus its polarization independence is limited to the power splitting ratio. 

\section{Symmetric PICs fabricated by FLM}

We present in this Section two novel methods to fabricate directional couplers whose splitting ratio is independent from light polarization and that produce polarization transformations that do not depend upon the specific input-output path.  We may denominate such a device as Symmetric Polarization Insensitive Coupler (SPIC).

All the devices discussed in the following are fabricated by FLM employing a Yb:KYW cavity-dumped femtosecond laser oscillator, with central emission wavelength at 1030 nm, pulse duration of $\sim$ 300 fs and repetition rate of 1 MHz. The substrate used is an alumino-borosilicate glass (Eagle XG, Corning Inc.). Waveguides  are inscribed at the depth of 170 $\mmu$ beneath the sample surface, and are optimized to yield single-mode operation at 1550 nm. All the optical characterizations discussed in the following are therefore performed at this wavelength.

\subsection*{Method 1: birefringence compensation}
\begin{figure*}[t]
\centering
\includegraphics[]{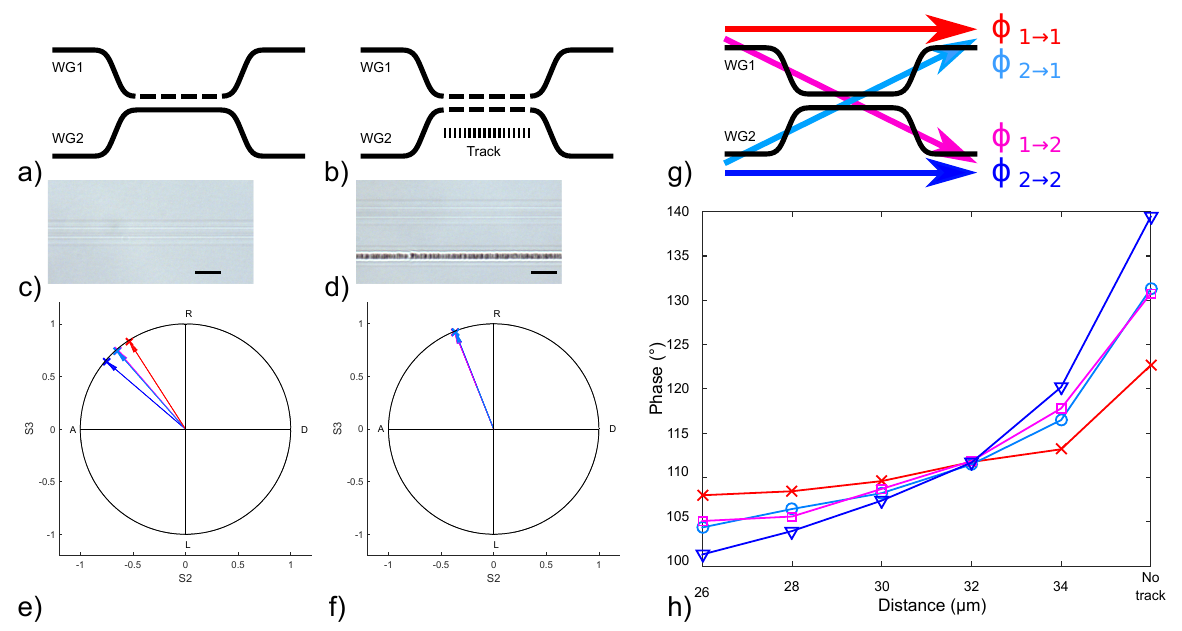}
\caption{(a) Schematic of a directional coupler fabricated by fs-laser micromachining. Inscription of the second waveguide modifies the birefringence of the first one, inscribed previously. (b) The addition of a damage track, with tuned irradiation parameters, next to such second inscribed waveguide, leads to birefringence equalization. (c),(d) Microscope pictures of the interaction region of the directional couplers without and with the inscription of the additional track (at distance $d_t$ = 32 $\mmu$). Scalebar is 20~$\mmu$. (e),(f) Stokes parameters of the output states for couplers without and with the additional track, for diagonally polarized input light and various input-output combinations. A planar projection of the Poincar\'{e} sphere is represented. (g) Light that enters in waveguide $m$ and exits from waveguide $n$ acquires a phase shift $\phi_{m \rightarrow n}$ between the horizontally and the vertically polarized components. (h) The different values of $\phi_{m \rightarrow n}$ are plotted as a function of $d_t$ (distance of the track from the second waveguide). For $d_t$~=~32~$\mmu$ the four possible $\phi_{m \rightarrow n}$ coincide.}
\label{f2}
\end{figure*}
The first method is based, and further elaborates, on the PIC concept that was demonstrated in Ref.~\cite{pitsios2017}. There, the  polarization insensitivity in the splitting ratio was produced by writing the two waveguides composing the couplers very close to each other. The inscription of the second waveguides alters the optical properties of the first one (Fig.~\ref{f2}a), affecting on the one hand the coupling coefficients $k_{H,V}$ for the two polarizations, and on the other hand the birefringence of the first waveguide \cite{pitsios2017,fernandes2012}. As a consequence, for a certain optimal waveguide separation, the beating frequencies for the two polarizations are experimentally observed to become equal and the power splitting ratio of the coupler can be made polarization-insensitive. However, since the birefringence of the two waveguides is unbalanced, the induced polarization rotations may be path-dependent as discussed in the previous section. Our idea is to inscribe with the femtosecond laser an extra track in the glass, in the proximity of the coupling region and next to the second waveguide, in such a way that the track produces on the second waveguide the same effect that the inscription of the second waveguide had produced on the first one, thus equalizing the birefringence of the waveguides of the coupler. 

We started our experimental investigation with the optimization of a planar PIC as the one demonstrated in Ref.~\cite{pitsios2017}.  Waveguide fabrication is operated with laser pulses of energy $E$~=~480~nJ and sample translation speed $v$~=~15 mm/s, adopting the optical configuration described in Ref.~\cite{corrielli2014} for the irradiation. In detail, a high-numerical aperture microscope objective (NA~=~1.42, oil immersion) is underfilled by focusing on its aperture the incoming laser beam, by means 50~cm focal length lens. The resulting waveguides exhibit propapagtion losses < 0.6 dB/cm. This method allows also the integration of waveguide-based waveplates with tilted axis. \cite{corrielli2014}.

Figure~\ref{f2}c shows a microscope picture of the interaction region of a coupler with waveguides separated by $d$~=~7$\mmu$ for a length $L$~=~300~$\mmu$, which yields a  difference in splitting ratio for H and V polarizations smaller than 1\%.  To characterize the polarization transformations we measured the normalized Stokes parameters at the output of each port of the device when diagonally-polarized light is injected in each input. Results are reported in Fig.~\ref{f2}e where it is shown the projection of the Poincar\'{e} sphere on the $S_2S_3$ plane. It is noted that all measured points lie, within the experimental uncertainty, on the equator $S_1 = 0$, indicating that light maintains a perfect degree of polarization and that no power exchange between H and V components has taken place in the coupler. 

The phase shift between H and V components can be retrieved as
\begin{equation}
\phi =\arctan \frac{S_3}{S_2} \label{angle}
\end{equation}
and corresponds in the graph to the angle between the horizontal axis ($S_2=0$) and the radius connecting the center to the measured equatorial point.
In agreement with our theoretical model, the phase shifts relative to the cross transfers (magenta and light blue crosses) are practically identical, $\phi_{1 \rightarrow 2}=131.3\gradi$, $\phi_{2 \rightarrow 1}=130.8\gradi$, and correspond to the quantity $\Phi$ of equations (\ref{Jones1}). As well, the bar transfer phases shift instead deviate from the cross ones by the same absolute amount, which corresponds to $\psi$ in \eqref{Jones1}, but opposite signs ($\phi_{1 \rightarrow 1}=122.7\gradi$, $\phi_{2 \rightarrow 2}=139.4\gradi$). 
This confirms the presence of a birefringence unbalance within the coupling region.

We also performed the same kind of measurement on a straight waveguide of the same length of the coupler. There, the phase shift was measured as $\phi_\mathrm{SWG}=158.0 \gradi$. Since $\phi_\mathrm{SWG}>\phi_{1 \rightarrow 2},\phi_{2 \rightarrow 1}$ and given the expression of the phase $\Phi$ obtained for a directional coupler (equation (\ref{relevant4})), we can infer that the effect of the second waveguide inscription is that of reducing the birefringence of the first one. This fact is further confirmed by the fact that we measured $\phi_{1,1}<\phi_{2,2}$ and thus a negative value of $\Psi$, which implies $b_1<b_2$ (see equation \ref{relevant5}). It is worth to highlight that a similar effect (with opposite sign) has been observed for laser written waveguides in fused silica \cite{fernandes2012}.

After this characterization, we fabricated a series of five directional couplers with exactly the same geometrical parameters as before, but with the addition of an extra damage track next to the interaction region, placed at a fixed distance $d_t$ from the second waveguide of the coupler and inscribed after it. The track flanks the coupler also for half of the curved waveguide segments, both at the fan-in and fan-out sides, always remaining at a distance $d_t$. The five couplers scan the value of $d_t$, ranging from 26 $\mmu$ to 34 $\mmu$ (fig.~\ref{f2}d shows a microscope picture of the interaction region of the coupler, in the case $d_t$~=~32~$\mmu$). In order to avoid parasitic light coupling in the damage track we strongly detuned it with respect to the waveguide; in detail, we inscribed it using the same irradiation power used for the waveguide writing, but we drastically reduced the translation speed to 0.5~mm/s.
We confirmed experimentally that the inscription of the tracks with these parameters did not alter the balanced PIC functioning and did not introduce additional optical loss.
	 
The values of the phase shifts between the two polarizations for such couplers are plotted as a function of $d_t$ in Fig.~\ref{f2}h, measured with diagonally-polarized input light. One note that for $d_t$ = 32 $\mmu$, the four points are overlapped, indicating that the polarization rotation is the same for each input-output combination. In fact, the inscription of the track causes a birefringence decrease for both waveguides of the coupler (as confirmed by the fact that $\phi_{1 \rightarrow 2}$ and $\phi_{2 \rightarrow 1}$ become smaller and smaller for shorter $d_t$). However, this reduction is stronger for the second waveguide, which is closer to the track: for $d_t$ = 32 $\mmu$ we obtain that the birefringence difference is completely compensated. Note that we cannot use the analytic model for further quantitative considerations because there we have not considered the coupling occurring in the curved waveguide segments present in a real directional coupler. Still, that model allows for a clear understanding of the observed phenomena. 
 
\begin{figure*}
\centering
\includegraphics[]{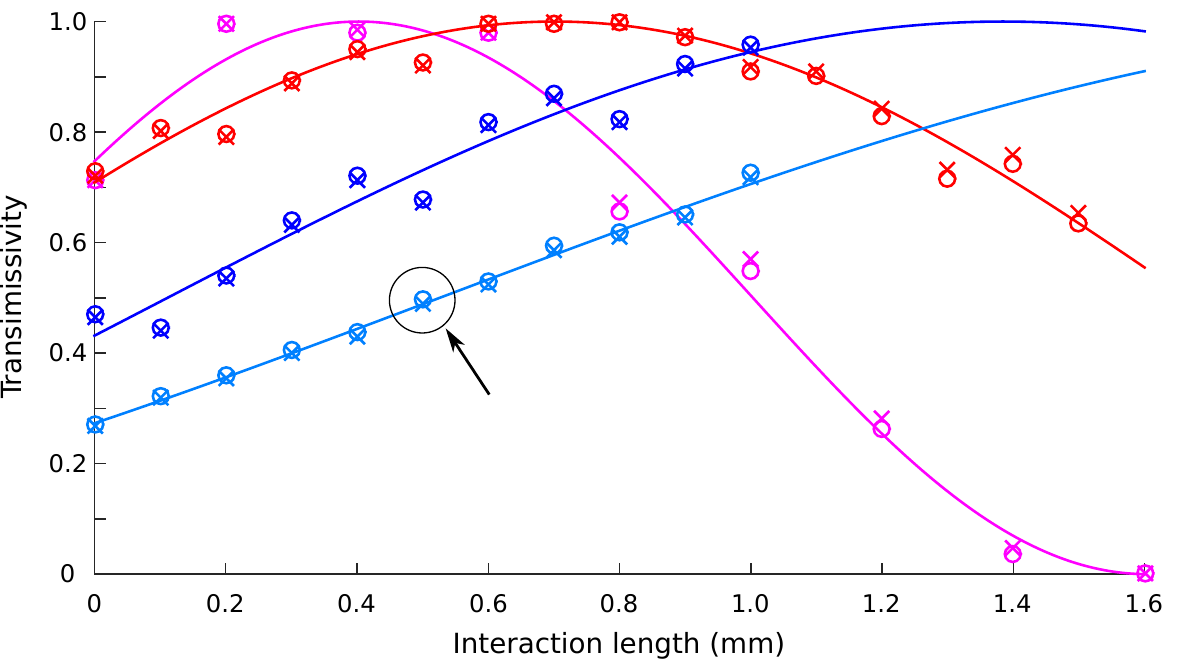}
\caption{Power transfer on the cross modes, as a function of the interaction length, for SPICs fabricated with the annealing technique. Different colors correspond to different coupling distances (magenta $d= 9 \mmu$, red $d= 10 \mmu$, blue $d= 11 \mmu$, light blue $d= 12 \mmu$). Measurements are performed both with horizontally (circles) and vertically (crosses) polarized input light. Continuous lines are best-fit sinusoidal functions. Further characterization of the polarization behaviour has been performed on the coupler marked with the black arrow.}
\label{f3}
\end{figure*}  
\subsection*{Method 2: thermal annealing}
The second method we present for the fabrication of a SPIC relies on laser written waveguides post-processed by thermal annealing. Arriola and co-workers \cite{arriola2013} recently demonstrated that it is possible to obtain single mode waveguides with enhanced confining and loss properties, by performing thermal annealing on former multimode waveguides inscribed with high pulse energy in borosilicate glass substrate. The annealing enables indeed a complex material relaxation process that both reduces the size of the guiding region, and increases the core/cladding refractive index contrast. Here we show that waveguides fabricated with this method also present a particularly low value of birefringence, and enables plainly the realization of SPICs.

We optimized the irradiation parameters to obtain (after annealing) low-loss single mode waveguides for the 1550~nm wavelength. We used a 0.6~NA  microscope objective as focusing optics and optimum conditions were found for pulse energy $E$~=~500~nJ and translation speed $v$~=~40~mm/s; each waveguides is fabricated by scanning the laser 10 times along the same path in the same translation direction. Before annealing such waveguides result to be highly multimode. For the thermal annealing we employed a procedure similar to the one reported in Ref.~\cite{arriola2013}: a first heating ramp at the rate of 100 $\gradi$C/h, from room temperature to 600~$\gradi$C, is followed by another ramp of 75~$\gradi$C/h, up to 750~$\gradi$C. Once the maximum temperature is reached, the sample is cooled down to room temperature at -12$\gradi$C/h rate. The overall duration of the process is about 70 hours. The annealed waveguides are single-mode at for 1550~nm wavelength, with  8.3~$\times$~8.9~$\mmu^2$ mode size ($1/e^2$). Measured propagation loss are $<$ 0.3~dB/cm.\\
The birefringence of these waveguides was characterized using the method described in Ref.~\cite{corrielli2014}, and we found a value of $b=1 \cdot 10^{-6}$, which is remarkably smaller ($\approx$20 times) than typical values measured for FLM written waveguides \cite{sansoni2010} in comparable irradiation conditions. We then fabricated 47 directional couplers with waveguides brought close at distances of 9~$\mmu$, 10~$\mmu$, 11~$\mmu$ and 12~$\mmu$, for interaction lengths ranging between 0 and 1.6~mm. 
The experimentally measured coupler transmissivities  $T$, defined as the fraction of light coupled to the cross mode with respect to the overall output light, are plotted in Fig.~\ref{f3}, for both horizontally and vertically polarized input light. No significant differences between the two polarizations are evidenced in the splitting ratio, independently from the geometrical characteristics of the couplers. The absolute difference in transmissivity between polarization, defined as $\epsilon=|T_H-T_V|$, is measured on average to be $\epsilon_{av}=0.7\%$, with a maximum value $\epsilon_{max}=2.1\%$. We characterized the polarization transformation for the coupler fabricated with $d$~=~11~$\mmu$ and $L$~=~0.5 mm (indicated by the arrow in Fig.~\ref{f3}), which yields $T_H=49.6\%$ and $T_V=49.0\%$. 
Also for this coupler we verified that the power transfer between the H and V polarized modes is negligible for all input output combinations. The measured phase shifts on the $S_2S_3$ plane are $\phi_{1 \rightarrow 1}=7.0\gradi$, $\phi_{1 \rightarrow 2}=6.8\gradi$, $\phi_{2 \rightarrow 1}=6.7\gradi$, $\phi_{2 \rightarrow 2}=6.6\gradi$, which are indeed very similar  one to each other. These results prove the validity of this method for realizing SPICs by femtosecond laser writing.

\section{Conclusions}

We have studied both theoretically and experimentally the behavior of PICs composed of birefringent waveguides. In particular, we have shown that a birefringence unbalance in the PIC interaction region leads to a polarization transformation performed by the coupler that is dependent from the specific path taken by the light within the device. We have then proposed and validated two different approaches that allow to eliminate this effect, based on femtosecond laser written waveguides. 

The two methods shown are complementary, in the sense that the first one, based on local birefringence compensation, easily allows to embed a SPIC in a birefringent optical circuit that may contain also polarization manipulation elements as polarizing beam splitters and optical waveplates\cite{crespi2011, corrielli2014, pitsios2017}. The second method, instead, is not suitable for a polarization manipulation platform development, but permits to realize a fully polarization insensitive optical circuit in a very simple and robust fashion. 

These results further widen the capabilities of femtosecond laser written circuits for integrated polarization manipulation. The SPICs developed in this work may be used as building blocks for large interferometric networks which are transparent to polarization, with possible applications in advanced quantum information experiments such as quantum walks or boson sampling, where polarization adds to path as a further available degree of freedom. In addition, a SPIC fabricated by local birefringence engineering has been recently employed in the development of an integrated source of polarization entangled photons \cite{atzeni2017}.

\acknowledgments
This work was supported by the ERC-Advanced Grant CAPABLE (Composite integrated photonic platform by femtosecond laser micromachining; grant agreement no. 742745).

\end{document}